\documentclass[12pt]{article}

\usepackage{amsmath}
\usepackage{amssymb}
\usepackage{natbib}

\usepackage{setspace}

\newcommand{\di}{\,{\rm div \,}}

\newcommand{\1}{{\bf 1}}

\newtheorem{theorem}{Theorem}
 
 \newtheorem{lemma}{Lemma}

 \newenvironment{proof}{\noindent{\bf Proof }}{\rule{2.5mm}{2.5mm}}

\newtheorem{xample}{Example}

\newcommand{\ep}{\varepsilon}

\newcommand{\R}{{\mathbb R}}

\newcommand{\cF}{{\cal F}}

\begin{document}

\thispagestyle{empty}

\title{Asset Trading in Continuous Time: \\ A Cautionary Tale}

\author{William R. Zame\thanks{I have benefitted  from comments from   Robert Anderson, Peter Bossaerts, Bryan Ellickson, and a seminar audience at CalTech.  The hospitality of the Geneva Finance Research Institute and University College, London, and financial support from the National Science Foundation, the UCLA Academic Senate Committee on Research, and the Geneva Finance Research Institute are gratefully acknowledged.  Views expressed here are those of the author and do not necessary reflect the views of any funding agency.} \\
{\parbox{3.5in}{\begin{center}
{\small Department of Economics\\ University of California, Los Angeles \\ \bigskip}
First Version: December 2001
\end{center}}} } 

\date{This Version: March 2022}

\thispagestyle{empty}

\maketitle

\thispagestyle{empty}
\vfill

\pagebreak

\begin{abstract}  The continuous time model of dynamic asset trading is the central model of modern finance. Because trading cannot in fact take place  at every moment of time, it would seem desirable to show that the continuous time model can be viewed as the limit of models in which trading can occur only at (many) discrete moments of time.  This paper demonstrates that, if we take terminal wealth constraints and self-financing constraints as seriously in the discrete model as in the continuous model, then the continuous trading model need not be the limit of discrete trading models.  This raises serious foundational questions about the continuous time model.

\medskip


\medskip


\end{abstract}
\thispagestyle{empty}

\pagebreak

\setcounter{page}{1}

\setstretch{1.5}

\section{Introduction }\label{sect:intro}  

The continuous time model of dynamic asset trading introduced by  \cite{Merton(1969)} is the central model of modern finance.  It embodies the idea that information arrives frequently in very small increments and that asset trading can occur frequently.  It allows application of the powerful techniques of stochastic processes and stochastic differential equations.  It is the setting for the  \cite{Black&Scholes(1973)} formula for the pricing of options and other derivative securities, and for the rigorous demonstration by \cite{Duffie&Huang(1985)} and 
\cite{Anderson&Raimondo(2008)} of the suggestion of \cite{Kreps(1982)} that frequent trading of long-lived assets can provide the equivalent of complete markets.

Of course, trading cannot be literally continuous, but must take place at discrete moments in time, If we wish to be assured that models in which trading can be continuous are good idealizations of a world in which trading must take place at discrete moments in time, it would seem natural to insist that models in which trading can be continuous are limits -- in some appropriate sense -- of models in which trading must be discrete.  This paper suggests, to the contrary, that models in which trading can be continuous {\em need not be} the limits of models in which trading must discrete.  This seems to raise some serious foundational questions about continuous trading models.

At first glance there might seem to be little difficulty in approximating continuous trading by discrete trading.  To be definite, consider a simple setting: information is revealed gradually from time $0$ to time $T$, consumption takes place only at time 0 and time $T$, a finite set of assets (yielding dividends only at the terminal time) is available for trade (at  given prices) at every moment in time.  Now consider a consumer who holds some initial endowment  $e$ at the terminal date $T$ and who wishes to achieve some terminal consumption $x$.  To this end, the consumer can  exercise some continuous trading strategy $\theta$ to achieve the desired net trade $z= x - e$.   This trading strategy may  require some  investment at date $0$ but should not require any investment thereafter; i.e., 
it should be {\em self-financing}.  The most obvious way to approximate $\theta$ by a discrete trading strategy is to choose  some finite sequence of times $0 = t_0 < t_1 < \ldots < t_k = T$,  define a strategy $\widehat{\theta}$ that trades only at the times $t_i$ (for $i = 0, \ldots, t_{k-1}$) and is constant on the intervening time intervals $(t_i, t_{i+1}]$ (so that $\widehat{\theta}(t) = \theta(t_i)$ for each $t \in (t_i, t_{i+1}]$) and show that, as the intervals between successive times shrink to $0$,  the discrete strategies 
$\widehat{\theta}$ converge (in some sense) to the continuous strategy $\theta$ and that 
the net trades  that these strategies  support converge (in some satisfactory sense) to the desired net trade $z$.  

However, even if the trading strategies and  supported net trades converge (in some satisfactory sense), this would not seem to be enough.  The problem is that the  net trades $\widehat{z}$ that are supported by the discretized trading strategies 
$\widehat{\theta}$  need not be {\em feasible} --  the bundles 
$\widehat{z} + e$ need not be non-negative -- so that the trading strategies $\widehat{\theta}$ could not actually be executed.  Thus it would seem that we should look for some {\em other} discrete trading strategies  
$\tilde{\theta}$ that do achieve feasible net trades $\tilde{z}$ and for which the achieved net trades 
$\tilde{z}$ converge to the net trade $z$.    Theorem 1 of this  paper shows that {\em such strategies need not exist}, and Example 1 shows that  non-existence can be a robust phenomenon.  Theorem 2 shows that non-existence may persist, even if we enlarge the class of discrete trading strategies to allow for strategies that trade infinitely often -- faster and faster as the terminal time approaches -- and Example 2 shows, again, that  non-existence can be a robust phenomenon.  (As the reader will see, both of these Examples are built in a very familiar framework that encompasses the standard Black-Scholes story, so there is nothing ``peculiar'' about them.) 

These  conclusions   are quite different from those reached in a body of work in the finance literature.   \cite{Duffie&Protter(1992)}, and Bertsimas, Kogan and Lo (2000), for example, demonstrate conditions under which convergence of price processes and trading strategies guarantees convergence of the associated gains process; \cite{Duffie&Protter(1992)} argues in an example  that continuous trading strategies can be approximated by discrete trading strategies.  However, these papers -- like most of the finance literature -- do not treat {\em equilibrium models} and -- perhaps as a consequence -- do not take terminal wealth constraints seriously.  In particular, in constructing discrete trading strategies that approximate a given continuous trading strategy, they do not require that the discrete trading strategies satisfy terminal wealth constraints, even when the continuous trading strategies they are designed to approximate are required to satisfy these same terminal wealth constraints.   Our results show that requiring that the discrete trading strategies satisfy terminal wealth constraints and self-financing constraints imposes severe restrictions, and that these restrictions may not be satisfiable, even in relatively simple settings.

The present paper is not the only one to raise questions about the connection between discrete and continuous time modeling.  Budish, Cramton and Shim (2015) and \cite{BudishCramtonKyle} 
(the closest papers of which I am aware) argue that, although it might be reasonable to model the flow of information as continuous (as we do here), it is not reasonable to model asset trading as continuous.   In a different vein, \cite{SadzikStacchetti} demonstrate that the connection between the continuous time principal-agent model of \cite{Sannikov} and discrete-time approximations of that model is quite sensitive to the exact specification of the information structure in the discrete-time approximations.

The results of this paper suggest that  the continuous trading model does not rest on secure economic foundations; finding such foundations would seem an important direction for research.  The Conclusion suggests some thoughts toward this end.

\section{Framework } \label{sect:framework}

Here we set out the framework in which we set the problem.  Because our motivation comes from thinking about equilibrium, we discuss equilibrium in both the static and dynamic settings.  We follow  \cite{Duffie&Huang(1985)} closely, and refer the reader there for further detail.  \cite{Karatzas&Shreve(1974)}, \cite{Jacod(1979)} and  \cite{Duffie(1988)} are good sources for background on stochastic processes and stochastic integration and connections to economics and finance.

\subsection{The Static Economy }  The economy evolves over a finite interval $[0,T]$.  A single good is available for consumption at the initial date $0$ and the terminal date $T$; there is no consumption at intermediate dates $t \in (0,T)$.   Uncertainty about the state of nature at date $T$ is modeled as a probability space $(\Omega, {\cal F}, P)$.  A date $T$ commodity bundle is a (measurable) function $x: \Omega \to \R$; we restrict to date $T$ commodity bundles that have finite variance.  Hence the commodity space is ${\mathcal X} = \R \times L^2 = \R \times L^2(\Omega, {\cal F}, P)$.  We write  
$\widehat{x} = (x_0,x) \in \R \times L^2$ for a typical  bundle; $x_0 \in \R$ is certain consumption at  time 0 and $x \in L^2$ is random consumption at time $T$.  We write $\1_F$ for the characteristic function of $F \in {\cal F}$.

There are $I$ consumers, each described by a consumption set, taken to be the positive cone $\R_+ \times L^2_+$; an endowment $\hat{e}^i = (e_0^i, e^i) \in \R_+ \times L^2_+$, and a utility function $U^i$.  Throughout, we assume that endowments are bounded away from $0$.  In the Examples, we will assume that utility functions $U^i$ have the form:
$$
U^i(x_0,x) = u^i(x_0) + \int_\Omega u^i(x(\omega)) \, dW(\omega)
$$
where $u^i : \R \rightarrow \R$ is some strictly increasing, strictly concave felicity function.

A {\em Walrasian (Arrow-Debreu) equilibrium} is a pair $\langle \widehat{p}, (\widehat{x}^i) \rangle $ consisting of a price $\widehat{p} = (p_0,p)\in \R_{++} \times L^2_{++}$ and an $I$-tuple of consumption choices 
\hbox{$\widehat{x}^i \in \R _+ \times L^2_+$} such that
\begin{itemize}
\item[(i)] for each $i$: $\widehat{p} \cdot x^i \leq \widehat{p} \cdot \widehat{e}^i$
\item[(ii)] for each $i$: if $\widehat{y}^i \in \R_+ \times L^2_+$ and $U^i(\widehat{y}^i) > U^i(\widehat{x}^i)$ then 
$\widehat{p} \cdot \widehat{y}^i > \widehat{p} \cdot \widehat{x}^i$
\item[(iii)] $\sum_i \widehat{x}^i = \sum_i \widehat{e}^i$
\end{itemize}

\subsection{The Dynamic Economy} In the dynamic economy, information is revealed gradually over time and assets can be traded as information is revealed.

\subsubsection{The Evolution of Information }  We describe the revelation of information by a {\em filtration} 
$(\cF_t)$, an increasing family of sub-$\sigma$-algebras of $\cF$; the $\sigma$-algebra $\cF_t$ represents information available at time $t$.  We assume that the filtration $(\cF_t)$ satisfies the usual conditions:
\begin{itemize}
\item $\cF_0$ is the $\sigma$-algebra of null sets (nothing is known at time 0)
\item $\cF_T = \cF$ (everything is known at time $T$)
\item $\cF_t \subset \cF_{t'}$ if $t < t'$
\item $\bigcap \{\cF_{t'} : t < t' \} = \cF_t$ for each $t$
\end{itemize}
In our Examples, the information filtration will be that generated by a Standard Brownian motion.  In that case we  could take $\Omega$ to be $C_0[0,T]$, the space of continuous real-valued functions that vanish at $0$, take $\cal F$ to be $\sigma$-algebra generated by the cylinder sets, take 
${\cal F}_t$ take be the sub-sigma algebra generated by cylinder sets in $C_0[0,t]$, and take $P$ to be Wiener measure.

As usual, it is convenient to view  a stochastic process either as a function  $S: [0,T] \times \Omega \to \R$ or as a time-indexed family of random variables on $\Omega$; in the latter case we write $S(t) = S(t, \cdot): \Omega \to \R$.  

\subsubsection{Assets, Asset Prices, and Trading Strategies}

An {\em asset} is a claim to consumption at the terminal time.  Asset $A$ yields {\em dividends} $A(\omega)$ at the terminal time if the state of the world is $\omega$; assets yield no intermediate dividends.   We assume throughout that asset dividends are bounded below, so there is some $C > 0$ for which $A(\omega) \geq -C$ for each state $\omega$.  (We allow asset dividends to be bounded above so that our framework includes the standard Black-Scholes model.)

$J+1$ assets $A_0, \ldots, A_J$ are available for trade.  (In the Examples we assume, as is done in many financial models, that the asset $A_0$  is riskless ($A_0 \equiv 1$) and the other assets are risky.)  A {\em portfolio} is a tuple $\varphi = (\varphi_0, \ldots, \varphi_J) \in \R^{J+1}$; $\varphi_j$ is the holding of asset $j$. The {\em dividend} on the portfolio $\varphi$ is the function $\di \varphi : \Omega \to \R$ defined by
$$
\di \varphi (\omega) = \sum_j \varphi_j A_j (\omega)
$$

A {\em trading strategy} is a stochastic process 
$$
\theta = (\theta_0, \ldots, \theta_J) : \Omega \times [0,T] \rightarrow \R^{J+1}
$$
We interpret $\theta(t,\omega)$ as the portfolio of assets held by a consumer at time $t$ if the state of the world is 
$\omega$.  A {\em price process} for the asset $A$ is a semi-martingale $S$ on the probability space 
$(\Omega, {\cal F}, P)$ that is adapted to the information filtration $(\cF_t)$.  
Given  price processes $S = (S_0, \ldots, S_J)$ for the assets $A_0, \ldots, A_J$ and a trading strategy $\theta$ we write $  \theta(t)  \cdot S(t)$ for the random value/cost of 
$\theta$ at time $t$.

The trading strategy $\theta$ is {\em admissible} if:
\begin{itemize}
\item[(i)] $\theta$ is {\em predictable} (roughly speaking, $\theta(t)$ depends only on information available 
{\em before} time $t$) 
\item[(ii)] each $\theta_j$ is square integrable with respect to the quadratic variation process $[S_j]$ for $S_j$ 
\item[(iii)] the gains process
$$
G = \int \theta \, dS = \sum_j \int \theta_j \, dS_j
$$
is well defined as a stochastic integral
\end{itemize}
The admissible trading strategy $\theta$ is {\em self-financing} if
$$
\theta(t)  \cdot S(t)   = \theta(0)  \cdot S(0)  + \int_0^t \theta \, dS 
$$
for every $t \in [0,T]$; that is, the current value of a portfolio is its initial value plus any gains or losses accumulated from trading.  The admissible trading strategy $\theta^i$ {\em supports} the consumption bundle 
$\widehat{x}^i$ if
\begin{itemize}
\item[(i)] $e^i_0 = \theta(0) \cdot S(0) + x^i_0$ 
\item[(ii)] $e^i +\di \theta(T) = x^i$
\end{itemize}

\subsubsection{Conditional Expectations}  For any integrable function $f: \Omega \to \R$ and and probability measure $Q$ on $\Omega$ that is boundedly absolutely continuous with respect to $P$,  we define the {\em conditional expectation of $f$ with respect to ${\mathcal F}_t$} $E_Q(f | {\mathcal F}_t)$ as the Radon-Nikodym derivative with respect to $Q$ of the measure $\mu$ on ${\mathcal F}_t$ defined by $\mu(X) = \int_X f \,dQ$ for each $X \in {\mathcal F}_t$. Thus, 
$$
\mu(X) =  \int_X f \,dQ = \int_X E_Q(f | {\mathcal F}_t) \, dQ
$$
for each $X \in {\mathcal F}_t$.  $E_Q(f | {\mathcal F}_t)$ is an ``averaged'' version of $f$.
Keep in mind that $E_Q(f | {\mathcal F}_t)$ is ${\mathcal F}_t$-measurable but $f$ may not be.  

\subsubsection{Risk-neutral Pricing} For simplicity, we assume that the information structure, the assets and the price processes satisfy  conditions that guarantee the existence of an equivalent {\em risk-neutral probability measure $Q$}.  That is
\begin{itemize}
\item[(i)] $P,Q$ are mutually absolutely continuous with bounded Radon-Nikodym derivatives.  In particular, $L^2(P) = L^2(Q)$.
\item[(ii)] For every time $t$, $S_j(t)\cdot A_j(t) = E_Q(A_j | {\cF_t})$
\end{itemize}
The existence of the risk-neutral probability measure implies that an admissible trading strategy 
$\theta$ is self-financing if and only if 
$$
E_Q(\di \theta(t) | {\cF_t}) = E_Q(\di \theta(t') | {\cF_{t}})
$$
for every $t, t' \in [0,T]$.

\subsubsection{Dynamic Equilibrium}

Given consumers $(e^i, u^i)$ and assets $A_0, \ldots, A_J$, a {\em dynamic (Radner) equilibrium} consists of  price processes $S$ and self-financing, admissible trading strategies $\theta^i$ and consumption plans $\widehat{x}^i$ for each consumer such that
\begin{itemize}
\item[(i)] for each $i$: $\theta^i$  supports $\widehat{x}^i$
\item[(ii)] for each $i$: if $\widehat{y}^i \in \R \times L^2_+$ and $U^i(\widehat{y}^i) > U^i(\widehat{x}^i)$  then 
$\widehat{y}^i$ cannot be supported by any self-financing, admissible trading strategy
\item[(iii)] $\sum_i \theta^i = 0$
\item[(iv)] $\sum_i \widehat{x}^i = \sum_i \widehat{e}^i$
\end{itemize}

\section{Simple Strategies}\label{sect:simple}  

Throughout this Section, we fix  the time interval $[0,T]$, the probability space
$(\Omega, {\cal F}, P)$, the filtration $(\cF_t)$ and  assets $A_0, \ldots, A_J$.   (Prices play no role here so we do not specify them.)  Fix a date $T$ endowment $e \in L^2_+$, assumed to be bounded away from $0$, and a date $T$ target consumption  $x \in L_+^2$ (which might be consumption in some unspecified equilibrium); write $z = x-e$ for the target net trade.  

Following  \cite{Harrison&Pliska(1981)}, we say the trading strategy $\theta$ is {\em simple} if there are a finite sequence  of times $0 = t_0 < t_1 < \ldots < t_N = T$ and bounded, ${\cal F}_{t_i}$-measurable functions $h_{ij}$ such that $\theta(t) = (h_{i0}, \ldots, h_{iJ}) $
for each $t \in (t_i, t_{i+1}]$.  (That is, $\theta$ trades only at the times $t_0, \ldots, t_{N-1}$, and $\theta(t)$ depends only on information available {\em before} time $t$.)   

We say that the simple strategy $\theta$  {\em respects the terminal wealth constraint} if 
$\di \theta(T) + e \geq 0$ (i.e., the terminal dividends yield a feasible net trade).\footnote{Note that this condition does not depend on prices.}  The simple strategy $\sigma$ {\em supports the net trade $z$} if $\di \theta(T) + e = z$.  By definition,  the sequence (or net) $(z^n)$ {\em converges weakly} to $z$ if
$$
\big\langle (z^n - z), f \big\rangle = \int_\Omega (z^n-z)(\omega) f(\omega) \, dP(\omega) \to 0
$$
for every $f \in L^2$.\footnote{We allow for nets, not just sequences, because the weak topology on $L^2$ is not metrizable.} 

The following Theorem derives strong necessary conditions for the existence of a sequence (or net) of simple strategies that respect the terminal wealth constraint and support net trades that converge weakly to the target net trade $z$.  Example 1, which follows, shows that the failure of these necessary conditions -- and hence the non-existence of such a sequence (or net) of simple strategies -- can be robust.

\begin{theorem}\label{thm:terminal} 
If there exists a sequence (or net) $(\theta^n)$ of simple strategies that respect the terminal wealth constraint and support net trades $(z^n)$ converging weakly to the net trade $z$ then:
\begin{itemize}

\item[{}] For every $\ep > 0$ there are a time $t^* < T$, a finite 
${\cal F}_{t^*}$-measurable partition $(X_\ell)$ of $\Omega$ and coefficients  
$\{\alpha_{j\ell}\}$ such that
\begin{itemize}
\item[(a)] $\sum \left( \alpha_{j\ell} \1_{X_\ell} \, A_j  \right) + e \geq  0$
\item[(b)] $\Vert z - \sum \left( \alpha_{j\ell} \1_{X_\ell} \, A_j \right) \Vert < \ep$
\end{itemize}

\end{itemize}
\end{theorem}
We defer the proof to  Appendix A.

We now  turn to the construction of an example that illustrates robust failure of the necessary conditions in Theorem 1.   Because part of the motivation for insisting that strategies respect terminal wealth constraints comes from thinking about equilibrium, we begin with the Walrasian equilibrium of a static, two-consumer exchange economy.  We then  consider a dynamic setting in which asset markets are (at least potentially) {\em dynamically complete}, so that the dynamic (Radner) equilibrium in which assets can be traded continuously replicates the consumptions in the Walrasian equilibrium of the static economy.  Because we will use the same framework for  Example 2 (which illustrates robust failure of the necessary conditions in Theorem 2), we present all the details and analysis of the static economy here.  

Let $B$ be a standard Brownian motion $B$ on the time interval $[0,T]$, with $(\Omega, {\mathcal F}, P)$ the underlying probability space.  Let $({\mathcal F}_t)$ be the filtration generated by $B$. Endowments and assets will be specified  as smooth functions of the terminal value of the Brownian motion $B$.  To define endowments,  choose any smooth function  $f: (-\infty, + \infty) \to (1, 3)$ with the following properties:
\begin{itemize}
\item $f'(c) > 0$  for $c \in (-\infty,0)$ and  $f'(c) < 0$ for $c \in (0,\infty)$
\item $f(0) > 3$
\item $f$ has limits: $\lim_{c \to -\infty} f(c) =  \lim_{c \to +\infty} f(c) = 1$
\item $f$ satisfies: $\gamma = \int_\Omega f(B(T,\omega)) \, dP(\omega) < 2$

\end{itemize}

We consider a two-consumer exchange economy.  Endowments are:
\begin{align*}
(e^1_0, e^1) = \big(1, f(B(T))\big) \, ; \, (e^2_0, e^2) = \big(3, 4 - f(B(T)) \big)
\end{align*}
Consumers $1,2$  maximize expected utility using the same smooth, strictly concave felicity function $u$ (the particular form is not important):
$$
U(x_0, x) = u(x_0) + \int_\Omega u(x(\omega)) \, dP(\omega)
$$

This economy has a unique Walrasian equilibrium; equilibrium consumptions are
$$
x^1 = (a, a); x^2 = (4-a, 4-a) 
$$ 
where $a < 2$ is a constant.\footnote{The value of $a$ depends on the felicity function $u$ and on the specific choice of $f$.}  To see this, note first that there is no aggregate risk in the economy; because both agents are strictly risk-averse this implies they smooth consumption perfectly at the terminal date.  Hence equilibrium consumptions are $x^1 = (a^1_0, a); x^2 = (4-a^1_0, 4-a)$.  At equilibrium, relative marginal rates of substitution are equal so 
$$
\frac{u'(a)}{u'(a^1_0)} = \frac{u'(4-a)}{u'(4-a^1_0)} 
$$ 
If $a > a^1_0$ then $4-a < 4-a^1_0$; because $u$ is strictly concave, this implies that $u'(a) < u'(a^1_0)$ and 
$u'(4-a) > u'(4-a^1_0)$, which would be a contradiction. Similarly, if $a < a^1_0$ then $4-a > 4-a^1_0$, which would again lead to a contradiction.  Hence $a^1_0 = a$ and $4 -a^1_0 = 4-a$, so equilibrium consumptions are 
$x^1 = (a, a); x^2 = (4-a, 4-a) $, as asserted.

Because date $1$ equilibrium consumptions are constant and agree with date $0$ consumptions, the date $0$ price of any date $1$ bundle coincides with its expectation (with respect to $P$).  Consumer 1's date $0$ endowment is strictly less than that of consumer 2 and, because $\gamma < 2$,  the date $0$ expectation of consumer 1's date $1$ endowment  is also strictly less than that of consumer 2.  Hence consumer 1's wealth must be strictly less than that of consumer 2, and the same must be true of their equilibrium consumptions; i.e., $a < 4-a$ and so $a < 2$ as asserted. 

With this analysis in hand we turn to the first illustrative example.

\medskip

\noindent {\bf Example 1 } There are two assets.  $A_0$ is riskless (i.e., $A_0 \equiv 1$).  $A_1$ is risky; its dividends are strictly positive and bounded above.  (In Example 2, the risky asset will be strictly positive but unbounded above.)   To define $A_1$, we fix a  smooth strictly increasing function $g: (-\infty,+\infty) \to (0,1)$  for which
$$
\lim_{c\to -\infty} g(c) = 0 \ \ \mbox{and} \ \lim_{c\to +\infty} g(c) = 1
$$
and define the asset $A_1$ by $A_1(\omega) = g(B(T,\omega))$ for each $\omega \in \Omega$. If $g$ is real analytic, it follows from \cite{Anderson&Raimondo(2008)}  that the asset market is dynamically complete; in particular, the Walrasian equilibrium can be implemented by continuous trading of the assets 
$A_0, A_1$.\footnote{\cite{Duffie&Huang(1985)} provide conditions on $A_1$ under which the asset market is dynamically complete, but it is not obvious what assumptions on $g$ would yield the requisite conditions on $A_1$.}

We claim that, no matter what functions $f, g$ we choose (subject to the given conditions),  {\em there do not exist simple trading strategies that respect consumer 1's terminal wealth constraint and support net trades approximating the Walrasian net trade $z$}.  To do this, we will show that there is some $\varepsilon > 0$ for which the conditions (a), (b)  of Theorem 1 do not obtain.\footnote{The $\varepsilon$ we construct depends only the function $f$, which defines the endowment of consumer 1, and the number $a$, which defines the consumption -- and hence the net trade -- of consumer 1, but not on the function $g$, which defines the risky asset.}  
 
Consider the Walrasian net trade $z = a - f(B(T))$ of consumer 1.  We have shown that $a < 2$;  by assumption, $f(0) > 3$, so $a - f(0) < - (1 + \mu)$ for some $\mu > 0$.  The terminal value of Brownian motion is normally distributed, so the set 
           $$
           F = \{ \omega \in \Omega : a - f(B(T,\omega)) < - (1 + \mu) \}
           $$
        has strictly positive measure: $P(F) > 0$.          
        Now fix a time $t^* < T$, a ${\mathcal F}_{t^*}$-measurable partition $\{X_\ell\}$ of 
 $\Omega$ and coefficients $\{ \alpha_{0\ell} \}, \{ \alpha_{1\ell} \}$.  Set 
 $$
 \Psi = \sum_\ell \Big[ \alpha_{0\ell} \1_{X_\ell} A_0 + \alpha_{1\ell} \1_{X_\ell} A_1 \Big]
 $$
and assume that $\Psi + e^1 \geq 0$.  We will show that $\Psi \geq -1 $ and then use that fact to show that 
$\Vert z - \Psi \Vert \geq \mu P(F)$. 

To show that $\Psi \geq - 1$ we begin by establishing this inequality on each set of the given partition.  To this end, fix a set $X_\ell$ in the given partition.  To show that  $\Psi(\omega) \geq - 1$ for $\omega \in X_\ell$, we first show that  \vspace{-12pt}
$$ 
    1 + \alpha_{0\ell} + \alpha_{1\ell} A_1(\omega) \geq 0  \vspace{-12pt}
 $$
    To accomplish this, fix $\delta > 0$.  The constraints we have imposed on $f, g$ imply that we can choose  $M$ so large that 
    \begin{alignat*}{2} \vspace{-12pt}
    x > +M \Rightarrow &\ 1 < f(x) < 1 + \delta & \mbox{ and } & 1- \delta < g(x) < 1 \nonumber\\
    x < - M \Rightarrow &\ 1 < f(x) < 1 + \delta & \mbox{ and } & 0 < g(x) < \delta  
    \end{alignat*} 
     Because  the increments of Brownian motion are independently and normally distributed, we can find $\cF$-measurable subsets $Y^-, Y^+ \subset X_\ell$ such that 
     \begin{alignat}{2}
    \omega\in Y^- \Rightarrow &\ B(T,\omega) < - M  \nonumber \\
    \omega\in Y^+ \Rightarrow &\ \ B(T,\omega) > + M  \label{eqn:basic}
    \end{alignat} 
     By definition, for $\omega \in X_\ell$ we have $\Psi(\omega) = \alpha_{0\ell} A_0(\omega) + \alpha_{1\ell} A_1(\omega)$, so if we substitute we find that  if 
     $\omega \in Y^-$ then
     \begin{align*}
   \big[ \Psi(\omega) + e^1(\omega) \big] - \big[1 + \alpha_{0_\ell} \big]  
                 &=     \big[f(\omega) + \alpha_{0\ell}  + \alpha_{1\ell} g(\omega) \big] 
                         - \big[1 + \alpha_{0_\ell} \big]  
                    < \delta + \delta  \alpha_1 
      \end{align*}
      and re-arranging yields
       \begin{align}
        \big[1 + \alpha_{0_\ell} \big]  &> 
            \big[\Psi(\omega) + e^1(\omega) \big] - \delta - \delta  \alpha_{1\ell}  \label{eqn:rearrange1}
       \end{align}
Similarly, if $\omega \in Y^+$ then
 \begin{align*}
     \big[ \Psi(\omega) + e^1(\omega) \big] - \big[1 + \alpha_{0\ell} + \alpha_{1\ell} \big] 
           &=  \big[f(\omega) + \alpha_{0\ell}  + \alpha_{1\ell} g(\omega) \big] \\
                   & \ \ \ \ \ \ \ - \big[1 + \alpha_{0\ell} + \alpha_{1\ell} \big]   \\
                      &< \delta + \delta  \alpha_{1\ell} 
      \end{align*}
      and re-arranging yields
       \begin{align}
        \big[1 + \alpha_{0\ell} + \alpha_{1\ell} \big]  
        &> \big[\Psi(\omega) + e^1(\omega) \big] - \delta - \delta  \alpha_{1\ell}  \label{eqn:rearrange2}
       \end{align}
       By assumption, $\Psi(\omega) + e^1(\omega) \geq 0$ so if  we let $\delta \to 0$ in the re-arrangements (\ref{eqn:rearrange1}) and (\ref{eqn:rearrange2}), we obtain
     \begin{align}  
 1 + \alpha_{0\ell} + \alpha_{1\ell} &\geq 0  \nonumber\\
    1 + \alpha_{0\ell} & \geq 0   \label{eqn:positivity}
    \end{align}
    
    By construction, $0 \leq A_1 \leq 1$ so $1 + \alpha_{0\ell} + \alpha_{1\ell} A_1(\omega)$ is a convex combination of $ 1 + \alpha_{0\ell} + \alpha_{1\ell} $ and $1 + \alpha_{0\ell}$ \,:
      \begin{align*} 
      1 + \alpha_{0\ell} + \alpha_{1\ell} A_1(\omega) &= A_1(\omega)[1 + \alpha_{0\ell} + \alpha_{1\ell}] + [1-A_1(\omega)][1 + \alpha_{0\ell}]
      \end{align*}
      In view of (\ref{eqn:positivity}), we conclude that $1 + \alpha_{0\ell} + \alpha_{1\ell} A_1(\omega) \geq 0$, for every $\omega \in X_\ell$, as desired. 
    
   We  now show that  $\Psi(\omega) \geq -1$ for every $\omega \in \Omega$.  Because $\{X_\ell\}$ is a partition of $\Omega$, we see that for every $\omega \in \Omega$
       \begin{align*} 
    1 + \Psi(\omega) &= 1 + \sum_\ell \1_{X_\ell}\Big[ \alpha_{0\ell} + \alpha_{1k} A_1(\omega) \Big] \\
      &= \ \sum_\ell \1_{X_\ell}\Big[1 + \alpha_{0\ell} + \alpha_{1\ell} A_1(\omega) \Big] \\
      & \geq \ 0 \  
    \end{align*}
    and hence that  $\Psi(\omega) \geq -1$    for every $\omega \in \Omega$, as desired.

    Finally, we use the fact that $\Psi \geq -1 $ to show that $ \Vert \Psi - z \Vert  \geq \mu P(F)$.
           By definition, $z = a - e = a - f(B(T,\omega))$.  Hence, for every $\omega \in F$ we have $z(\omega) < - (1 + \mu)$ and $(\Psi - z)(\omega) > \mu$.  Thus,
      \begin{align*}
        \Vert \Psi - z \Vert &= \left(\int_X  \big\vert (\Psi - z)(\omega) \big\vert^2 \, dP(\omega) \right)^{1/2} \\
                                                    &\geq \left(\int_F  \big\vert (\Psi - z)(\omega) \big\vert^2 \, dP(\omega)  \right)^{1/2}\\
                                           &\geq  \left( \mu^2 P(F)  \right)^{1/2} \\ 
                                         &\geq \mu P(F)
           \end{align*}
            which is the desired lower bound.
            
           We conclude that if $0 < \ep <   \mu P(F)$ then there cannot exist a time $t^* < T$, a partition  
            $\{X_\ell\}$ and coefficients $\{\alpha_{0\ell} \}, \{ \alpha_{1\ell} \}$ satisfying conditions (a), (b) of Theorem 1.  Hence, there cannot exist a sequence (or net) of simple strategies that respect consumer 1's terminal wealth constraint and support net trades that converge weakly to consumer 1's Walrasian net trade. This completes Example 1. $\Box$\
            
            To understand the bite of the terminal wealth constraints, begin by considering the  implications in equation (1).  The increments of Brownian motion between the time $t^*$ and the terminal time $T$ are normally distributed with mean $0$ and standard deviation $T-t^*$.  Hence if $T-t^*$ is small, then the differences
            $\vert B(\omega,T) -  B(\omega,t^*) \vert$ will be small for {\em most} $\omega \in X_\ell$.  Hence, if $M$ is large than the sets $Y^-, Y^+$ will be small; i.e., have low probability.  If were to allow the terminal wealth constraints to be violated with low probability, then  the inequalities (\ref{eqn:rearrange1}) and (\ref{eqn:rearrange2}) would also be violated with low probability, and so we would no longer obtain  the inequalities (\ref{eqn:positivity}).  These inequalities obtain precisely {\em because} we insist that the terminal wealth constraints be satisfied everywhere and  reject the possibility of  low probability violations.  We will return to this point in the Conclusion.

\section{Almost Simple Strategies }

The arguments for Theorem 1 and Example 1 rely on the fact that any simple trading strategy involves a last trade; the arguments could not be carried through if there were no last trade.  However, in this Section, we show that, even if there is no last trade, similar reasoning leads to similar (negative) conclusions, provided we assume that strategies are self-financing.  The reason is that there is a gap between {\em any two successive trading opportunities};  the requirement that the trading strategy be self-financing links all trades together and gives the terminal wealth constraint the same ``bite''  as it has for simple trading strategies.
 
  To formalize this idea, we borrow the terminology of  \cite{Harrison&Kreps(1979)} and  say the trading strategy 
  $\theta$ is {\em almost simple} if there are a (possibly infinite) sequence  of times $ 0 = t_0 < t_1 < \ldots <  T$  and ${\cal F}_{t_i}$-measurable functions $h_{ij} \in L^2$ such that $\theta(t) = (h_{i0}, \ldots, h_{iJ}) $
for each $t \in (t_i, t_{i+1}]$.
 Thus, an almost simple strategy trades only finitely many times in every interval $[0,t)$ for $t < T$ but perhaps trades infinitely many times in every interval $(t,T)$ -- trading faster and faster near the terminal date.  
 
We consider the same setting as in Theorem 1, except that, because we employ the self-financing constraint, we need to specify the price process as well.  To do so, we assume the existence of risk-neutral probability measure $Q$ which is mutually absolutely continuous with  $P$, and for which  the Radon-Nikodym derivative $dQ/dP$ is bounded above and bounded away from $0$.   Thus,  the price of the portfolio $\varphi$ at  time $t \leq T$ will be $E_Q(\di \varphi | {\mathcal F}_t)$ -- the expected dividends of the portfolio
 $\varphi$, taken with  respect to $Q$, conditional on information available at time $t$.  
 
  As before, we fix a date $T$ endowment $e \in L^2_+$ (assumed to be bounded away from $0$), a date $T$ target consumption  $x \in L_+^2$, and write $z = x-e$ for the target net trade.

 \begin{theorem}\label{thm:self-financing}
 If there exists a sequence (or net) $(\theta^n)$ of self-financing, almost simple trading strategies that support feasible net trades  $(z^n)$ converging weakly to the target net trade $z$, then: 
 \begin{itemize}
 \item[{}]  For every time $t^* \in (0,T)$ and $\ep > 0$ there are a time $t^{**} \in (0 , t^*)$,  a finite 
 $\cF_t^{**}$-measurable partition   $\{X_\ell\}$ of $\Omega$ and  coefficients  
$\{\alpha_{j\ell}\}$ such that
\begin{itemize}
\item[(a)] $E_Q\Big[  \big(\sum_{j,\ell} \alpha_{j\ell} \1_{X_\ell} A_j \big) + e   \,  \big| {\cal F}_{t^{*}} \Big]  \geq  0$

\item[(b)] $\Big\Vert E_Q\Big[  \big(\sum_{j,\ell} \alpha_{j\ell}  \1_{X_\ell} A_j  \big) - z  \, \big| {\cal F}_{t^{*}} \Big] \Big\Vert < \ep$
 \end{itemize}
\end{itemize}
\end{theorem}

We again defer the proof to  Appendix A.  The argument  parallels the argument for Theorem 1, with expectations 
$E_Q (\, \cdot \, | {\mathcal F}_{t*})$ playing the role of terminal consumptions.

We now  turn to the construction of an example that illustrates robust failure of the necessary conditions in Theorem 2.  We  use the same framework as used in Example 1, except that we use an unbounded risky asset and we need to specify the risk-neutral probability measure.  

\medskip

\noindent {\bf Example 2 } 
We treat the two-consumer exchange economy introduced before Example 1.  As before, we take $A_0$ to be the riskless asset, but for illustrative purposes we use a different risky asset.  Let $g: (-\infty,+\infty) \to (-\infty,+\infty)$ be any smooth, strictly increasing, convex function which has limits
$$
\lim_{x \to -\infty} g(x) = 0 \ \ \mbox{and } \ \lim_{x \to +\infty} g(x) = +\infty 
$$
and for which
\begin{align}
\int_{-\infty}^{+\infty} g(x)^2 \phi_{0,T}(x)  \,dx < \infty  \label{eqn:normA1}
\end{align}
Here, and below, we have written $\phi_{\mu,\sigma}$ for the pdf of the normal distribution with mean $\mu$ and standard deviation 
$\sigma$; we write $\Phi_{\mu,\sigma}$ for the corresponding cdf.  We define the risky asset $A_1$ by 
$$
A_1(\omega) = g(B(T,w))
$$
for each $\omega \in \Omega$.  The inequality (\ref{eqn:normA1}) guarantees that $\Vert A \Vert < \infty$ so $A \in L^2$.  (As before, if $g$ is real analytic -- in particular if $g$ is the exponential function -- then it follows from \cite{Anderson&Raimondo(2008)}  that the asset market will be dynamically complete; in particular, the Walrasian equilibrium can be implemented by continuous trading of the assets $A_0, A_1$. \cite{Duffie&Huang(1985)} provide other conditions on $A_1$ under which the asset market is dynamically complete, but it is not obvious what assumptions on $g$ would yield the requisite conditions on $A_1$.)   Because  equilibrium consumptions are riskless, we can take the risk-neutral probability measure $Q = P$.  

Before going further, it is convenient to  collect some facts about expectations of the functions $f$ and $g$; we defer the proofs to Appendix B.

\begin{lemma}
For every $\ep > 0$ there is a $t(\ep) < T$ such that if $t \in (t(\ep) ,T]$ and $x_0 \in \R$ then \vspace{-18pt}
\begin{align*}
\Big\vert E\big[ \,f(B(T) \, | \, B(t,\omega) = x_0 \, \big] - f(x_0) \Big\vert &< \ep \\
 \ E\big[ \, f(B(T) \, | \, B(t,\omega)) \leq x_0 \, \big] & <  f(x_0) + \ep
\end{align*}
\end{lemma}

\begin{lemma} For every $t < T$ and $c \in (-\infty,+\infty)$ \vspace{-12pt}
\begin{align*}
E\big[ \, g(B(T)) \, | \, g(B(t,\omega))\, \big] \geq c \, \big] \geq c
\end{align*}
\end{lemma}  

\begin{lemma}  For every $\ep > 0$ there exists $c(\ep)$ such that if $c \leq c(\ep)$ and $t \in [\ep,T)$ then \vspace{-18pt}
\begin{align*}
E\big[ \, g(B(T) \, | \, g(B(t,\omega))= c \, \big] & <   \ep \\
 \ E\big[ \, g(B(T) \, | \, g(B(t,\omega)) \leq c \, \big] & <   \ep
\end{align*}
\end{lemma}

As before,  we consider the Walrasian net trade $z = a - f(B(T))$ of consumer 1.  We have shown that $a < 2$;  by assumption, $f(0) > 3$, so $a - f(0) < - (1 + 2\mu)$ for some $\mu > 0$.  Because $f$ is strictly increasing on the interval $(-\infty,0)$ and decreasing on the interval $(0,+\infty)$, there are unique $\lambda_1 \in (-\infty,0), \lambda_2 \in (0,+\infty)$ for which $f(\lambda_1) = f(\lambda_2) = a + 1 + \mu$; equivalently, $a - f(\lambda_) = f(\lambda_2) = - (1 + \mu)$.  For each $t \in (0,T]$ define    
$$
F_t = \big\{\omega \in \Omega: \lambda_1 < B(T,\omega) < \lambda_2 \big\} 
           = \big\{ \omega \in \Omega : a - f(B(T,\omega))  < - (1 + \mu) \big\} 
$$
 Note that $ P(F_t) = \Phi_{0,t}(\sigma_2) - \Phi_{0,t}(\sigma_1)$.    
Because the  values of Brownian motion at time $t$ are normally distributed with mean 0 and standard deviation $t$, $P(F_t) > 0$ for each $t$; in particular, $P(F_T) > 0$.  Set $\ep = (\mu/4) P(F_T)$ and define $t(\ep)$ as in Lemma 1.  Because $P(F_t)$ is a continuous function of $t$,  we can choose 
$t^* \in \big(t(\ep),T \big) $ so that $P(F_{t^*}) > P(F_T)/2$.   
        
      Now, suppose for the purpose of obtaining a contradiction, that there are a time $t^{**}$, a $\cF_{t^{**}}$-measurable partition 
      $\{X_\ell\}$ of $\Omega$ and coefficients 
      $\{\alpha_{0\ell} \}, \{\alpha_{1\ell}\}$ satisfying conditions (a), (b) of Theorem 2.  Set 
    \begin{align*}
      \Psi = \sum_{\ell} \big[ \left( \alpha_{0\ell}A_0 + \alpha_{1\ell}A_1 \right) \1_{X_\ell} \big]
   \end{align*}
      We will show that $\Vert E[ \, \Psi - z \, | \, \cF_{t^*} \, ] \Vert > \ep$, which will be contradiction.      
      As in Example 1, we proceed by first obtaining a lower bound on $\Psi$.  The argument parallels that in Example 1; however, because we need to worry about expectations (conditional on $\cF_{t^*}$) rather than on terminal values, some fussiness is required.  
      
     Fix an index $\ell$.  We first show that $\alpha_{1\ell} \geq 0$.  To see this suppose not, so that $\alpha_{1\ell} < 0$.  
     Fix $M > 0$.  The set
      $$
      Y_\ell(M) = \{ \omega \in X_\ell : g(B(t^*,\omega) \geq M \}
      $$ 
      is $\cF_{t^*}$-measurable and, because the increments of Brownian motion are normally distributed and $g$ is increasing, it has positive measure.   Condition (a) tells us that       $E[ \Psi + e \, | \, \cF_{t^*} ] \geq 0$, so we must have
      $$
      E[ \alpha_{0\ell} + \alpha_{1\ell} A_1 + e \, | \, \cF_{t^*} ](\omega) \geq 0
      $$
      for every $\omega \in Y_\ell(M)$.  By definition, $A_1(\omega) = g(B(T,\omega)   \geq 0$. Lemma 2  implies that  $E[ g(T,\omega) \, | \, Y_\ell ] \geq M$ for every $\omega \in Y_\ell(M)$.  
      We have supposed that $\alpha_{1\ell} < 0$, so
$$
      0 \leq E[ \alpha_{0\ell} + \alpha_{1\ell} A_1 + e \, | \, \cF_{t^*} ](\omega) <  E[ \alpha_{0\ell}  + e \, | \, \cF_{t^*} ](\omega)  + \alpha_{1\ell}M
      $$
      for every $\omega \in Y_\ell(M)$ 
      However the first term on the right-hand side is fixed and -- because $\alpha_{1\ell} < 0$ -- we can make the second term arbitrarily negative. This would yield a contradiction, so we conclude that  $\alpha_{1\ell} \geq 0$, as desired.
      
      We now show that $\alpha_{0\ell} \geq  - 1 - \ep$.  To see this, fix $m > 0$ and define
       $$
      Y'_\ell(m) = \{ \omega \in X_\ell : g(B(t^*,\omega) \leq m \}
      $$
     $Y'_\ell(m)$ is $\cF_{t^*}$-measurable and, because the increments of Brownian motion are normally distributed, it has positive measure.   As before, the requirement (a) that $E[ \Psi + e \, | \, \cF_{t^*} ] \geq 0$ implies that 
      $$
      E[ \alpha_{0\ell} + \alpha_{1\ell} A_1 + e \, | \, \cF_{t^*} ](\omega) \geq 0
      $$
      for every $\omega \in Y'_\ell(M)$.  For every $\delta > 0$, Lemma 3 implies that if we choose $m$ sufficiently small then $E[g(B(T)) \, | \, Y'_\ell ](\omega) < m(1+\delta)$ for every  $\omega \in Y'_\ell(M)$.  Hence, for $\omega \in Y'_\ell(M)$ we must have 
      $$
      0 \leq E[\Psi + e \, | \,\cF_{t^*}] = E[ e + \alpha_{0\ell} + \alpha_{1\ell}A_1 | \,\cF_{t^*}]  
             \leq (1 + \ep) + \alpha_{0\ell} + \alpha_{1\ell} m(1+\delta)\
      $$
If we let $m \downarrow 0 $ we obtain  $1 + \ep + \alpha_{0\ell} \geq 0$ and   $\alpha_{0\ell} \geq -1 - \ep$, as desired.

We have already seen that $\alpha_1 \geq 0$ and we have assumed $A_1 \geq 0$ so  
$$
E[\Psi(\omega) \, | \,\cF_{t^*}](\omega) = \alpha_{0\ell} + \alpha_{1_\ell} E[A_1| \,\cF_{t^*}](\omega) \geq -1 - \ep
$$
   for every $\omega \in X_\ell$.  Because $\{X_\ell\}$ is a partition of $\Omega$, we conclude that
   $$
E[\Psi \, | \,\cF_{t^*}](\omega)  \geq -1 - \ep
$$
for every $\omega \in \Omega$.  
   
 Lemma 1 implies that 
 $$
\Big| E[\, f(B(T) \,| \,\cF_{t^*} \,](\omega) - f(B(t^*,\omega)) \, \Big| < \ep
 $$
By definition, $e = f(B(T))$ so 
$$
\Big| E[\, e \, | \, \cF_{t^*} \,](\omega) - f(B(t^*,\omega) \, \Big| < \ep
$$  
Because  $z = a - e^1$, it follows that, if $\omega \in F_{t^*}$ then 
\begin{align*}
 E[\, \Psi - z \, | \, \cF_{t^*} \,](\omega) &=
             E[\, \Psi  \, | \, \cF_{t^*} \,](\omega) - E[\, a   \, | \, \cF_{t^*} \,](\omega) + E[\, e^1  \, | \, \cF_{t^*} \,](\omega) \\
             &\geq -(1+\mu) - \ep - a + f(B(t^*,w)) - \ep \\
             &\geq -(1+\mu) - \ep + (1+2\mu) -  \ep \\ 
             &= \mu - 2\ep   
              \end{align*} 
 By construction, $P(F_{t^*}) > P(F_T)/2$.   It follows that 
 $$
 \Big\Vert E[\, \Psi - z \, | \, \cF_{t^*} \,] \Big\Vert^2 \geq (\mu - 2\ep) P(F_{t^*}) \geq (\mu/2)P(F_{t^*}) \geq (\mu/4)P(F_T) = \ep
 $$
 Hence, for the chosen $\ep$ and constructed $t^*$,  there cannot exist a time $t^{**} < T$, a partition  
            $\{X_\ell\}$ and scalars $\{\alpha_{0\ell} \}, \{ \alpha_{1\ell} \}$ satisfying conditions (a), (b) of Theorem 2.  Hence there cannot exist a sequence (or net) of almost simple strategies that respect consumer 1's terminal wealth constraint and support net trades that converge weakly to consumer 1's Walrasian net trade.  $\Box$

\section{Conclusion }\label{sect:conclusion}

This paper argues that, if we take terminal wealth constraints seriously, then it is not obvious to how to interpret continuous trading as a limit of discrete trading.  This poses a challenge for further research.  Several possible approaches come to mind.  

Because the negative conclusions obtained here are driven by the requirement that terminal wealth must be non-negative, it might be tempting to adopt a model in which terminal wealth is allowed to be arbitrarily negative.   However, such a model would seem problematic:  What does negative consumption mean?  How is it possible for the consumption of  a single consumer to exceed the social endowment (as would be possible if the consumption of some other consumer were negative)?  Leaving aside these (and perhaps other) difficulties of interpretation, allowing arbitrarily negative consumption would certainly mean abandoning many of the utility functions (constant relative risk aversion, for instance) most commonly used -- especially in finance.  

A second approach might be to discretize, not only trading opportunities, but also information arrival, asset dividends, and endowments, and thus to view the continuous time model as a limit of discrete time models with {\em different} information structures, {\em different} asset dividends, and {\em different} endowments -- as well as {\em different} trading opportunities.  This is the approach followed by \cite{CoxRoss&Rubinstein(1979)} (which derives the Black-Scholes option pricing formula as a limit of binomial option pricing formulae) and by \cite{He(1990)} (which connects equilibrium in discrete-time models with equilibrium in continuous time models) and it is the approach followed in all computational models (of which I am aware).  But this approach also seems somewhat problematical.  Like the Black-Scholes formula itself, the binomial option pricing formula rests on the possibility of synthesizing an option by dynamic trading, and such synthesis rests on the possibility of trading the underlying asset {\em whenever its price moves}; if the price of the underlying asset can {\em move} every millisecond, it must be possible to {\em trade} every millisecond.  Otherwise, we must confront the difficulty that the strategy which exactly replicates an option when trades can be made every millisecond may not be executable at all when trades can be made only every other millisecond.

A more attractive alternative might be to adopt a model in which default is possible and might occur at equilibrium.  Such a model has been used by \cite{Zame(1993)} to demonstrate (in a different environment) that default can play an important role in substituting for missing asset markets.  However, the conclusions of such a model might well depend on the particular way in which default and the consequences of default are modeled.   \cite{Zame(1993)} follows (Dubey, Geanakoplos and Shubik (2005)  in modeling the consequences of default in terms of utility penalties, while  \cite{Geanakoplos&Zame(2014)} models the consequences of default as the seizure of collateral -- and these papers reach rather different conclusions about the implications of default. This suggests that the implications of incorporating default in the model might depend -- perhaps quite sensitively -- on the particular model of default chosen.  Moreover, as all three of the cited papers show, if default is possible then it will often be optimal for agents to  {\em choose} trading strategies that lead to default (with positive probability) -- even in the presence of a complete set of state-contingent asset markets.  

A quite different -- and more speculative -- alternative is to think about noise in the model(s), as suggested by the game-theoretic work of  \cite{LevinePesendorfer(1995)}  They show that equilibrium in the usual model of a game with a continuum of players can be very different from equilibrium in games with a large finite number of players.  In the former model, individual agents cannot be distinguished and individual deviations cannot be observed or punished; in the latter model, even ``very small'' agents can be distinguished and individual deviations can be observed and punished.  \cite{LevinePesendorfer(1995)} shows that the addition of (arbitrarily small amounts of) noise to the model(s) restores a close connection between the equilibria of the two models.

\nocite{BertsimasKogan&Lo(2000)}
\nocite{DubeyGeanakoplosShubik}
\nocite{He(1990)}

\nocite{Geanakoplos&Zame(2014)}

\nocite{BudishCramtonShim}

\bibliographystyle{econ}

\bibliography{CT-2021}

\pagebreak

\section*{Appendix A: Proofs of Theorems 1 and 2}

\begin{proof}{\bf of Theorem 1 } 
Let  $\mathbb{S}$ be the set of simple strategies.  A convex combination of simple strategies is a simple strategy, so $\mathbb{S}$ is a convex subset of the space of all strategies.  Define  $D: \mathbb{S} \to L^2$ by $D(\sigma) = \di \sigma(T)$.  $D$ is a linear map so $D(\mathbb{S})$ is a convex subset of $L^2$. $L^2_+$ is a convex subset of $L^2$, so $L^2_+ - e$ and $D(\mathbb{S}) \cap (L^2_+ - e)$ are also convex subsets of $L^2$.  Mackey's theorem   (see \cite{Aliprantis&Border(1999)}  for example) guarantees that the weak and norm closures of $D(\mathbb{S}) \cap (L^2_+ - e)$ coincide.  By assumption, the target net trade $z$ belongs to the weak closure of $D(\mathbb{S}) \cap(L^2_+ - e)$ so it also belongs to the norm closure.  Hence, given  
$\delta > 0$ (to  be chosen later), we can choose   a simple strategy $\theta$ such that $\Vert \di \theta - z \Vert < \delta$ and $\di \theta + e \geq 0$. We will show that if $\delta$ is small enough then we can obtain (a) and (b).

We begin by choosing $\rho$ sufficiently close to $1$ that $\Vert (1-\rho) \di \theta(T) \Vert < \delta $
 and setting $\tilde{\theta} = \rho \theta$.  By assumption,  $e(\omega) \geq b$ for every $\omega \in \Omega$, so 
$$
(\di \tilde{\theta}(T) + e)(\omega) = (\rho \di \theta + e)(\omega) \geq (1-\rho) b > 0 
$$
for every $\omega \in \Omega$.  By construction, $\theta$ is a simple strategy, so $\tilde{\theta}$ is also a simple strategy.  Hence there is some time $t^* < T$ and there are bounded, ${\cal F}_{t^*}$-measurable functions $h_j$ such that $\tilde{\theta}(t^*) = (h_0, \ldots, h_J)$ for each $t \in [t^*,T]$.  In particular, 
$D(\tilde{\theta}) = \di \tilde{\theta}(T) = \sum h_j A_j$.  

By assumption, the functions $h_j$ are ${\cal F}_{t^*}$-measurable and bounded.  Choose and fix some $c >  \sup \{\vert h_j(\omega) \vert : 0 \leq j \leq J, \omega \in \Omega\}$.  
 Choose a grid of points $-c = c_0 < c_1 \ldots < c_K = +c$ with the property that $c_k - c_{k-1} < \delta$ for each .  For each $j = 0, 1, \ldots, J; k = 1, \ldots, K$, define $ Y_{jk} = h_j^{-1}\big( (c_{k-1}, K] \big)$.
 For fixed $j$, the sets $\{Y_{jk}\}$ form a ${\cal F}_{t^*}$-measurable partition of $\Omega$ and, for each $\omega \in Y_{jk}$ we have 
 $ c_k - \delta < c_{k-1} < h_j(\omega) \leq c_k$.   By assumption, asset dividends are bounded below; say 
 $A_j(\omega) \geq - C$ for all $k, \omega$.  Hence, if $\omega \in Y_{jk}$ we have
  \begin{align*}
 c_k A_j(\omega) - h_j(\omega) A_j(\omega) \geq - C\delta  \end{align*}
We can rewrite this as
\begin{align}
 c_k \1_{Y_{jk}}(\omega) A_j(\omega) - h_j(\omega) \1_{Y_{jk}} A_j(\omega)  \geq - C\delta  \label{eqn:kj}
\end{align}
 Note that, because both terms are $0$ for $\omega \not\in Y_{jk}$, the inequality (\ref{eqn:kj}) holds for all $\omega \in 
 \Omega$.  Moreover, because the sets $Y_{jk}$ are disjoint, we also have
 \begin{align}
 \sum_k \left[ c_k \1_{Y_{jk}}(\omega) A_j(\omega)\right] - \sum_k \left[h_j(\omega) \1_{Y_{jk}} A_j(\omega)\right]  \geq - C\delta  \label{eqn:sumkj}
 \end{align}
For each $j$, the family $\{Y_{jk}\}$ is a partition of $\Omega$ so the second term on the left side is just
\begin{align}
\sum_k \left[h_j(\omega) \1_{Y_{jk}} A_j(\omega)\right]  &= h_j(\omega) A_j(\omega)  \label{eqn:2ndterm}
\end{align}

Now define $\{X_\ell\}$ to be the least common refinement of the partitions  $\{Y_{jk}\}$.  For each $j,k$, define  
$L(j,k) = \{\ell: X_\ell \subset Y_{jk}\}$ and note that, by construction, the collection $ \{X_\ell : \ell \in L(j,k)\}$  is a partition of $Y_{jk}$.  Moreover, for every $\ell$ and every $j$, there is a {\em unique} $k$ such that $X_\ell \subset  Y_{jk}$; we  define $ \alpha_{j\ell} = c_k$  for that unique $k$.  Note that 
$\alpha_{j\ell} = \alpha_{j\ell'} = c_k$ whenever $\ell, \ell' \in L(j,k)$.  

We now show that, it $\delta$ is small enough then the partition $\{X_\ell\}$ and the coefficients $\{\alpha_{j\ell} \}$ satisfy conditions (a), (b).  
 
For given $j, \ell$, there is a unique $k$ with $X_\ell \subset Y_{jk}$ so $\alpha_{j\ell} = c_k$.  Hence
\begin{align*}
\alpha_{j\ell} \1_{X_\ell} A_j - h_j \1_{X_\ell} A_j  = (c_k - h_j)\1_{X_\ell} A_j \geq -C\delta
\end{align*}
Because $\{X_\ell\}$ is a partition of $\Omega$ it follows that
\begin{align*}
\sum_\ell \alpha_{j\ell} \1_{X_\ell} A_j -  h_j  A_j    \geq -C\delta
\end{align*}
and hence that 
\begin{align*}
\sum_{j,\ell} \alpha_{j\ell} \1_{X_\ell} A_j - \sum_j h_j  A_j \geq - (J+1)C\delta
\end{align*}
By construction, $ \sum_j h_j  A_j + e \geq (1-\rho)b$ so 
\begin{align*}
\sum_{j,\ell} \alpha_{j\ell} \1_{X_\ell} A_j + e &\geq  \sum_j h_j  A_j  + e - (J+1)C\delta \geq  (1-\rho)b - (J+1)C\delta 
\end{align*}
Thus, if $\delta$ is small enough then 
\begin{align*}
\sum_{j,\ell} \alpha_{j\ell} \1_{X_\ell} A_j + e &\geq  0
\end{align*}
which is (a).

To see (b),  use the triangle inequality, recall that $\Vert  \sum_j h_j A_j - z \Vert < \delta$ (by construction), and use the definition of $\Psi$ to obtain 
\begin{align*}
\left\Vert \sum_{j,\ell} \alpha_{j\ell} \1_{X_\ell} A_j - z \right\Vert &\leq \left\Vert \sum_{j,\ell} \alpha_{j\ell} \1_{X_\ell} A_j
                                                                        - \sum_j h_j A_j \right\Vert + \left\Vert   \sum_j h_j A_j - z \right\Vert \\
                             & \leq \left \Vert \sum_{j,\ell} (\alpha_{j\ell} -h_j) \1_{X_\ell} A_j \right\Vert + \delta \\
                             &= \left\Vert  \sum_j \left[\sum_\ell  (c_k  - h_j)\1_{X_\ell} \right] A_j \right\Vert + \delta \\
                             &\leq  \sum_j \delta \left\Vert  A_j \right\Vert + \delta \\
                             &= \delta \left(  \sum_j  \left\Vert  A_j \right\Vert + 1 \right)
\end{align*}
 The assets $A_j$ belong to $L^2$ so their norms are finite, so again, if $\delta$ is small enough we will have 
 $\Vert \sum_{j,\ell} \alpha_{j\ell} \1_{X_\ell} A_j \Vert  < \ep$, which is (b), so the proof is complete.   \end{proof}

\bigskip

\begin{proof}{\bf of Theorem 2}  The proof closely parallels that of Theorem 1, except that conditional expectations (with respect to the risk-neutral probability measure $Q$) at time $t^*$  play the role of consumptions at the terminal time $T$.

 Let  $\mathbb{A}$ be the set of almost simple  strategies.  A convex combination of simple strategies is a simple strategy, so $\mathbb{A}$ is a convex subset of the space of all strategies.  Define  $D: \mathbb{A} \to L^2$ by $D(\sigma) = \di \sigma(T)$.  $D$ is a linear map so $D(\mathbb{A})$ is a convex subset of $L^2$. $L^2_+$ is a convex subset of $L^2$, so $L^2_+ - e$ and $D(\mathbb{A}) \cap (L^2_+ - e)$ are also convex subsets of $L^2$.  Mackey's theorem  again guarantees that the weak and norm closures of $D(\mathbb{A}) \cap (L^2_+ - e)$ coincide.  By assumption, the target net trade $z$ belongs to the weak closure of $D(\mathbb{A}) \cap(L^2_+ - e)$ so it also belongs to the norm closure.  Hence, for each  $\delta > 0$ (to  be chosen later) there is an almost simple strategy $\theta$ such that $\Vert \di \theta(T) - z \Vert < \delta$ and $\di \theta(T) + e \geq 0$. 
 
Choose $\rho < 1$ so that 
$$
\Vert (1-\rho)\di \theta (T) \Vert < \delta
$$
 and set $\tilde{\theta} = \rho \theta$.  Because 
$\theta$ is almost simple, so is $\tilde{\theta}$.  Hence there  are a time $t^{**} < t^*$ and bounded, 
${\mathcal F}_{t^{**} }$- measurable functions $h_j$ such that $\tilde{\theta}(t) = (h_0, \ldots, h_J)$ for each $t \in [t^{**} , t^*]$.    Arguing exactly as in the proof of Theorem 1, we can find a  finite ${\mathcal F}_{t^{**} }$-measurable partition $\{X_\ell\}$ of $\Omega$ and scalars $\alpha{j\ell}$ such that $ 0 <  \beta_{j\ell}  - h_j(\omega) < \delta$ for each $j, \ell$ and each $\omega \in X_\ell$.  We now show that, it $\delta$ is small enough then the partition $\{X_\ell\}$ and the coefficients $\{\alpha_{j\ell} \}$ satisfy  (a), (b). 


To obtain (a), recall that, by construction, $\di \tilde{\theta}(t^*) + e \geq (1-\rho)b$.  Hence, because  $\tilde{\theta}$ is self-financing, we have
\begin{align*}
E_Q\left[ \sum_j  h_j A_j  + e \, | \, \cF_{t^*} \right] &= E_Q\left[ \di \tilde{\theta}(t^*) +e  | \, \cF_{t^*} \right]  \geq (1-\rho)b
\end{align*}
We have assumed that the assets $A_j$ are bounded below -- say $A_j \geq - C$ for each $j$ -- so their conditional expectations are also bounded below; $E_Q[A_j | \cF_t ] \geq - C$ for each $j$ and each time $t \leq T$.  Hence
\begin{align*}
\sum_{j,\ell} \alpha_{j\ell} \1_{X_\ell} A_j  -  \sum_j  h_j A_j  \geq -  (J+1) C \delta
\end{align*}
Putting these together yields
\begin{align*}
E_Q\left[ \sum_{j,\ell} \alpha_{j\ell} \1_{X_\ell} A_j  + e \, | \cF_{t^*} \right] 
                                              &\geq E_Q\left[ \sum_j  h_j A_j  + e \, | \, \cF_{t^*} \right] - (J+1)C\delta \\
                                              &\geq   (1-\rho)b -  (J+1)C\delta    
                                              \end{align*}
If $\delta$ is small enough then the last term is positive, so we have (a).

To see (b), note that first, for each asset $A_j$, Jensen's inequality yields
\begin{align*}
\Vert E_Q\left[ A_j | \cF_{t^*}\right] \Vert_Q^2 &= \int_\Omega \left( E_Q\left[ A_j | \cF_{t^*}\right] \right)^2 \, dQ  \\
                                &\leq \int_\Omega E_Q\left[ A_j^2 | \cF_{t^*}\right] \, dQ \\
                                &= \int_\Omega A_j^2  \, dQ    \\
                                &= \Vert A_j \Vert^2_Q
\end{align*}
where we have written $\Vert \cdot \Vert_Q$ for the norm in $L^2(\Omega, \cF, Q)$.  By assumption, $P, Q$ are mutually boundedly absolutely continuous, so $L^2(\Omega, \cF, Q) = L^2(\Omega, \cF, P)$ and there is some constant $K$ for which
\begin{align*}
\frac{1}{K} \Vert f \Vert_Q \leq  \Vert f \Vert_P \leq K \Vert f \Vert_Q
\end{align*}
for every $f \in L^2(\Omega, \cF, Q) = L^2(\Omega, \cF, P)$.  Thus 
\begin{align}
\Vert E_Q\left[ A_j | \cF_{t^*}\right] \Vert \leq K \Vert A_j \Vert  \label{eqn:norm}
\end{align}

Now use the triangle inequality, together with (\ref{eqn:norm}) and our choice of $\theta, \tilde{\theta}$ to obtain:

\begin{align*}
\left\Vert E_Q\left[ \sum_{j,\ell} \alpha_{j\ell} \1_{X_\ell} A_j  -z \, \Big| \cF_{t^*} \right] \right\Vert &=
        \left\Vert E_Q\left[ \sum_{j,\ell} \alpha_{j\ell} \1_{X_\ell} A_j  \, \Big| \cF_{t^*} \right]  - E_Q[ z  |  \cF_{t^*} ] \right\Vert \\
            & \leq  \left\Vert E_Q\left[ \sum_{j,\ell} \alpha_{j\ell} \1_{X_\ell} A_j  \, \Big| \cF_{t^*} \right] 
                     - E_Q\left[ \sum_j h_j A_j \Big| \cF_{t^*} \right ]\right\Vert \\
                     & \ \ \ \  + \left\Vert E_Q\left[ \sum_j h_j A_j  \Big| \cF_{t^*} \right]  - E_Q[ z  |  \cF_{t^*} ]\right\Vert \\
                     &= \left\Vert E_Q\left[ \sum_{j,\ell} (\alpha_{j\ell} - h_j)\1_{X_\ell} A_j  \, \Big| \cF_{t^*} \right]  \right\Vert \\ 
                      & \ \ \ \  + \left\Vert E_Q\left[ \sum_j h_j A_j  -z \Big| \cF_{t^*} \right]  \right\Vert  \\
             & \leq \delta \sum_j \Vert A_j \Vert_Q + \left\Vert E_Q\left[ \sum_j \di \tilde{\theta}(T)  -z \Big| \cF_{t^*} \right]  \right\Vert \\
             &\leq \delta K \sum_j \Vert A_j \Vert + K \left\Vert \sum_j \di \tilde{\theta}(T)  -z \right\Vert \\
             &\leq  \delta K \sum_j \Vert A_j \Vert  +  K \delta \\
             &= \delta \left( \sum_j \Vert A_j \Vert + K \right)
                      \end{align*}                                                                                      
If $\delta$ is small enough then the last term will be less than $\ep$, so we will have (b), and the proof is complete.  \end{proof}

\pagebreak

\section*{Appendix B: Proofs of Lemmas 1, 2 and 3 }  Recall that we write $\phi_{\mu,\sigma}$ and $\Phi_{\mu,\sigma}$ for the pdf and cdf (respectively) of the normal distribution with mean 0 and standard deviation $\sigma$.  We begin with Lemma 1, which addresses the function $f$.  Recall that we have assumed $f$ is smooth, increasing on $(-\infty,0)$, decreasing on $(0,\infty)$, and has  limits
$$
\lim_{x \to - \infty} f(x) =  \lim_{x \to + \infty} f(x) = 1
$$

\begin{proof}{\bf of Lemma 1 }  Given $\ep > 0$, choose $M$ so large that $f(x) <  1 + \ep/2$ for $x \in (-\infty,-M] \cup [+M, +\infty)$.  The interval $[-(1+M), +(1+M)]$ is compact, so $f$ is uniformly continuous on that interval. Hence there is some $\delta > 0$ such that if $x,x' \in [-(1+M), +(1+M)]$ and $\vert x - x' \vert < \delta$ then $\vert f(x) - f(x') \vert < \ep/2$; there is no loss in assuming that $\delta < 1$.  Because $f(x) <  1 + \ep/2$ for $x \in (-\infty,-M] \cup [+M, +\infty)$, it follows that if $x,x' \in \R$ and $\vert x - x' \vert < \delta$ then $\vert f(x) - f(x') \vert < \ep/2$.  Choose $t(\ep)$ sufficiently close to $T$ that $\Phi_{0,T-t(\ep)}(+\delta) - \Phi_{0,T-t(\ep)}(-\delta) > 1 - \ep/(2 f(0))$.  Note that if $t(\ep) \leq t  < T$ then we also have 
$\Phi_{0,T-t}(+\delta) - \Phi_{0,T-t}(-\delta) > 1 - \ep/(2 f(0))$.

Now fix $t$ with $t(\ep) \leq t  < T$ and $x_0 \in \R$.  Set $\sigma = T = t$.  Because the increments of Brownian motion are independent and normally distributed, we can write
\begin{align*}
E\big[ \,f(B(T) \, | \, B(t,\omega) = x_0 \, \big] &= \int_{-\infty}^{+\infty} f(x_0 + x) \phi_{0,\sigma}(x) \, dx
\end{align*}
We can decompose the integral in the following way:
\begin{align*}
\int_{-\infty}^{+\infty} f(x_0 + x) \phi_{0,\sigma}(x) \, dx &= \int_{-\infty}^{-\delta} f(x_0 + x) \phi_{0,\sigma}(x) \, dx \\ 
          & \ \ \ \ + \int_{-\delta}^{+\delta} f(x_0 + x) \phi_{0,\sigma}(x) \, dx \\
          & \ \ \ \ + \int_{+\delta}^{+\infty} f(x_0 + x) \phi_{0,\sigma}(x) \, dx 
          \end{align*}
          To estimate the first and third integrals, note that by construction, $1 \leq f(x) \leq f(0)$ for all $x$, and $\Phi_{0,\sigma}(+\delta) - \Phi_{0,\sigma}(-\delta) > 1 - \ep/(2 f(0))$, so that the sum of the first and third integrals is bounded by $f(0) [\ep/(2 f(0))]$.  To estimate the second integral, first use the triangle inequality to write
    \begin{align*}  
      \left\vert \int_{-\delta}^{+\delta} f(x_0 + x) \phi_{0,\sigma}(x) \, dx    - f(x_0) \right\vert 
            &\leq  \left\vert \int_{-\delta}^{+\delta} \big[ f(x_0 + x) - f(x_0) \big] \phi_{0,\sigma}(x) \, dx  \right\vert \\
                        & \ \ \ \ +  \left\vert \int_{-\delta}^{+\delta} f(x_0) \phi_{0,\sigma}(x) \, dx - f(x_0) \right\vert 
                                \end{align*}    
       By construction, $\vert f(x_0 + x) - f(x_0) \vert < \ep/2$ for $x \in (-\delta, + \delta)$ so the first term on the right-hand side is at most $\ep/2$; we have chosen $\sigma = T-t \leq T-t(\ep)$ so $1 - \ep/2f(0) < \Phi_{0,\sigma}(+ \delta) - \Phi_{0,\sigma}(- \delta) \leq 1$ so the second term on the right is less than $f(0)[\ep/2f(0)]$.  Putting this all together yields  
        \begin{align*} 
         \Big\vert E\big[ \,f(B(T) \, | \, B(t,\omega) = x_0 \, \big] - f(x_0) \Big\vert < 2 f(0)[\ep/2f(0)] = \ep
         \end{align*} 
         as desired.  The second inequality follows immediately.
          \end{proof}

We now turn to Lemmas 2 and 3, which involve the function $g$.  Recall that we have assumed that $g$ is smooth, strictly increasing,  convex, satisfies the limits
$$
\lim_{x \to - \infty} g(x) = 0 \ , \ \lim_{x \to + \infty} g(x) = +\infty
$$  
and has the property that
$$
\int_{-\infty}^{+\infty} g(x)^2 \phi_{0,1}(x) dx < \infty
$$

\begin{proof}{\bf of Lemma 2 } 
Because $g$ is strictly increasing it is invertible, so we can define $\mu = g^{-1}(c)$.  
We have assumed that $g$ is convex, so Jensen's inequality implies
\begin{align*}
E\big[ \, g(B(T)) \, | \, g(B(t,\omega)) \geq c \, \big] &= E\big[ \, g(B(T)) \, | \, B(t,\omega) \geq \mu \, \big] \\
                &\geq g\left( E\big[ \, B(T) \, | \, B(t,\omega) \geq \mu \, \big] \right) 
                \end{align*} 
The increments of Brownian motion are normally distributed with mean 0, so $E\big[\, B(T) \, | \, B(t,\omega) \geq \mu \, \big] \geq \mu$.  Keeping in mind that $g$ is increasing and $g(\mu) = c$, we have
\begin{align*}
E\big[ \, g(B(T)) \, | \, g(B(t,\omega)) \geq c \, \big] &\geq g\left( E\big[ \, B(T) \, | \, B(t,\omega) \geq \mu \, \big] \right) \geq g(\mu) = c
                \end{align*} 
which is the desired result. \end{proof}

\begin{proof}{\bf of Lemma 3 } To obtain the first inequality, we first compare the pdf's of the normal distributions with mean 0 and standard deviations $\sigma, T$.  By definition
\begin{align*}
\phi_{0,\sigma} &= \frac{1}{\sigma \sqrt{2\pi}} \exp\left[ - \left(\tfrac{1}{2}\right) \left(\tfrac{x}{\sigma}\right)^2 \right] \\
\phi_{0,T} &= \frac{1}{T \sqrt{2\pi}} \exp\left[ - \left(\tfrac{1}{2}\right) \left(\tfrac{x}{T}\right)^2 \right]
\end{align*}
A little algebra shows that we can choose $M_1$ sufficiently large that if $x \geq M_1$ and $\sigma \in (0,T-\ep]$ then
 \begin{align*}
\phi_{0,\sigma}(x) \leq \phi_{0,T}(c)
\end{align*}

By assumption, $\lim_{c\to -\infty} g(x) = 0$; hence, given $\ep > 0$ so there is some $m$ for which $g(m) < \ep/2$.  By assumption, $g$ is strictly increasing, so $g(x) < \ep/2$ for each $x \leq m$.  Note that
\begin{align*}
 \int_{-\infty}^{+\infty} g(x)^2 \phi_{0,T}(x) \,dx \geq \left[ \int_{-\infty}^{+\infty} g(x) \phi_{0,T}(x) \,dx \right]^2
\end{align*}
Hence we can find $M_2$  sufficiently large that 
\begin{align*}
 \int_{M_2}^{+\infty} g(x) \phi_{0,T}(x) \,dx < \ep/2
\end{align*}

Set $M = \max\{M_1, M_2\}$; without loss, we may assume $M > m$.  Define $c(\ep) = g(m - M)$; note that, because $g$ is increasing,  that  $c(\ep) < g(m) < \ep/2$.

Now fix $c \leq c(\ep)$ and set $\mu = g^{-1}(c)$.  Because $g$ and $g^{-1}$ are increasing, $\mu \leq m-M$.
Because the increments of Brownian motion from time $t$ to time $T$ are normally distributed with mean 0 and variance $\sigma^2 = T-t$, we have 
\begin{align*}
E\big[ \, g(B(T) \, | \, g(B(t,\omega))= c \, \big] &= \int_{-\infty}^{+\infty} g(\mu + x) \phi_{0,\sigma}(x) \,dx         
\end{align*}
Decompose the integral as follows:
\begin{align}
 \int_{-\infty}^{+\infty} g(\mu + x) \phi_{0,\sigma}(x) \,dx   &=  \int_{-\infty}^{M} g(\mu + x) \phi_{0,\sigma}(x) \,dx    \nonumber \\
                        & \ \ \ \ +  \int_{M}^{+\infty} g(\mu + x) \phi_{0,\sigma}(x) \,dx      \label{eqn:intg}
\end{align}
Because $\mu \leq m-M$, $g(\mu + x) \leq g(m) < \ep/2$ for every $x \in (-\infty, M]$ and the normal distribution has total mass 1,  the first integral on the right-hand side of (\ref{eqn:intg}) is at most $\ep/2$.  Because $g$ is increasing, $\mu \leq m-M < 0$ and 
$M = \max\{M_1, M_2\}$, we have
\begin{align*}
  \int_{M}^{+\infty} g(\mu + x) \phi_{0,\sigma}(x) \,dx  &\leq   \int_{M}^{+\infty} g(x) \phi_{0,\sigma}(x) \,dx \\
                       &\leq  \int_{M}^{+\infty} g(x) \phi_{0,T}(x) \,dx \\
                       &\leq  \int_{M_2}^{+\infty} g(x) \phi_{0,T}(x) \,dx 
 \end{align*}
Hence the second integral on the right-hand side of (\ref{eqn:intg}) is at most $\ep/2$.  Thus, 
\begin{align*}
E\big[ \, g(B(T) \, | \, g(B(t,\omega))= c \, \big] < \ep
 \end{align*}
 which is the first of the  inequalities asserted by the Lemma.  The second inequality is immediate, so the proof is complete.
 \end{proof}

\end{document}